\documentclass[twocolumn]{aastex62}
\pdfoutput=1 
\usepackage{amsmath,amstext}
\usepackage[T1]{fontenc}
\usepackage{apjfonts} 
\usepackage[figure,figure*]{hypcap}

\usepackage{amssymb}
\usepackage{natbib}

\def\rp{{$r$-process }}
\def\msun{{$M_\odot$}}

\shorttitle{Collapsars as an \rp Source}
\shortauthors{Macias \& Ramirez-Ruiz}

\begin{document}
\title{\Large \bf Constraining  Collapsar \lowercase{$r$}-Process  Models through Stellar Abundances}

\author[0000-0002-9946-4635]{Phillip Macias}
\author[0000-0003-2558-3102]{Enrico Ramirez-Ruiz}
\affiliation{Department of Astronomy and Astrophysics, University of California, Santa Cruz, CA 95064, USA}
\affiliation{DARK, Niels Bohr Institute, University of Copenhagen, Blegdamsvej 17, 2100 Copenhagen, Denmark}

\begin{abstract}

We use observations of heavy elements in very metal-poor stars ([Fe/H] < -2.5) in order to place constraints on the viability of collapsar models as a significant source of the \rp. We combine bipolar explosion nucleosynthesis calculations with recent disk calculations  to make predictions of the observational imprints these explosions would leave on very metal-poor stars. We find that a source of low ($\approx 0.1-0.5$ \msun) Fe mass which also yields a relatively high ($> 0.08$ \msun)  \rp mass would, after subsequently mixing and forming new stars, result in  [$r$/Fe] abundances up to three orders of magnitude higher than those seen in stars. In order to match inferred  abundances, 10-10$^3$ \msun\, of Fe would need be efficiently incorporated into the \rp ejecta. We show that Fe enhancement and hence [$r$/Fe] dilution from other nearby supernovae is not able to explain the observations unless significant inflow of pristine gas occurs before the ejecta are able to form new stars. Finally, we show that the inferred [Eu/Fe] abundances require levels of gas mixing which are in conflict with other properties of \rp enhanced metal-poor stars. Our results suggest that early \rp production is likely to be spatially uncorrelated with Fe production, a condition which can be satisfied by neutron star mergers due to their large kick velocities and purely \rp yields.

\end{abstract}

\section{Introduction}

The gravitational wave discovery of the binary neutron star merger (NSM) GW170817 \citep{abbott2017} along with the intensive multi-wavelength electromagnetic observations of the ensuing kilonova SSS17a/AT2017gfo \citep[e.g.,][]{coulter2017, cowperthwaite2017, drout2017, shappee2017} provided definitive evidence of NSMs being a viable source of \rp elements in the galaxy and universe as a whole. Since then, a number of interesting questions have arisen regarding the physics of NSMs as well as the implications of NSMs being the only significant source of the \rp throughout cosmic time \citep[see][for a summary of current issues with the merger-only model]{siegel2019}. 

One puzzle exhibited by the kilonova is the large mass of lanthanide rich ejecta ($\approx 0.035 $ \msun) inferred through modeling \citep[e.g.,][]{kasen2017}. While the mass of dynamical ejecta expelled by tidal forces is expected to be at most $\approx 10^{-2}$ \msun\, \citep[e.g.,][]{radice2018}, the kilonova required several times this mass in order to explain the peak luminosity and subsequent evolution. This requires an additional source of heavy \rp ejecta, with one theoretically predicted candidate being outflows from a remnant accretion disk. These outflows, which originally were thought to be powered by neutrino irradiation and thus result in a relatively weak \rp, may be instead launched by nuclear recombination and viscous dissipation, maintaining a low $Y_{\rm e}$ and resulting in a heavy \rp \citep[e.g.,][]{wu2016}. 

In addition, while numerous studies have implicated a rare (in comparison to standard core-collapse supernovae) channel for the \rp at both early and late times \citep{hotokezaka2015,ji2016, macias2018}, some issues have arisen with NSMs being a dominant channel in the context of galactic chemical evolution. In particular, the expected / observed  delay time distribution $t^{-1}$ of NSMs predicts a flat [Eu/Fe] trend in stars with [Fe/H] metallicities > -1, where we use the standard notation [El/H] $=  {\rm log_{10}(}N_{\rm \text{El}}/N_{\rm H}) -  {\rm log_{10}(}N_{\rm \text{El}}/N_{\rm H})_\odot$, where $N_{\rm El}$ is the column density of a given element. This is not seen in the data, as the [Eu/Fe] exhibits a ``knee" at around [Fe/H]=-1, similar to that of alpha elements produced in CCSNe. This discrepancy implies that there may be another early, possibly dominant, channel of \rp production that has yet to be directly detected \citep{cote2018}, or that the delay-time distribution of NSMs differs from that of Type Ia supernovae \citep[e.g.,][]{simonetti2019}.  

    One solution recently detailed by \citet{siegel2019b} is the collapsar model proposed by \citet{woosley1993}, in which the core of a massive ($\gtrsim$30 \msun) rotating star collapses to form a black hole. The resultant explosion does not unbind the entirety of the star, leaving behind material with sufficient angular momentum to form an accretion disk \citep{macfadyen1999,lee2006}. Due to the large ($> 10^{-3} M_\odot$ s$^{-1}$)  accretion rates achieved at early times, the midplane of the disk may attain high enough density to suppress positron creation, thus capturing electrons only and producing neutron rich material. The ejection of this material through a disk wind would then produce a strong \rp. This scenario, motivated by the inferred accretion disk winds of SSS17a, would result in significant heavy element production immediately following star formation events, thus reproducing the trend in [Eu/Fe] at late times. However, due to the large progenitor masses required, collapsars are inherently rare (even compared to NSMs) and must produce a large mass of \rp ejecta per event in order to be consistent with the average mass production-rate inferred for the Milky Way of $\sim 10^{-7}$ \msun\, yr$^{-1}$ \citep{bauswein2014}. 

Here we discuss the implications of such large \rp ejecta masses and compare to observations of stars that have been formed from gas that has been enriched by at most a few nucleosynthetic production events. In Section \ref{sec:obs}, we derive predictions for stellar abundance measurements within the collapsar scenario to compare to observations of very metal-poor stars and discuss mixing of the \rp elements. In Section \ref{sec:discussion} we comment on the implications of our findings in the context of \rp progenitors.

\section{An Observational Constraint on the Collapsar Model}\label{sec:obs}

\subsection{Comparison with Theoretical Models}

\begin{figure*}[!ht]

\plotone{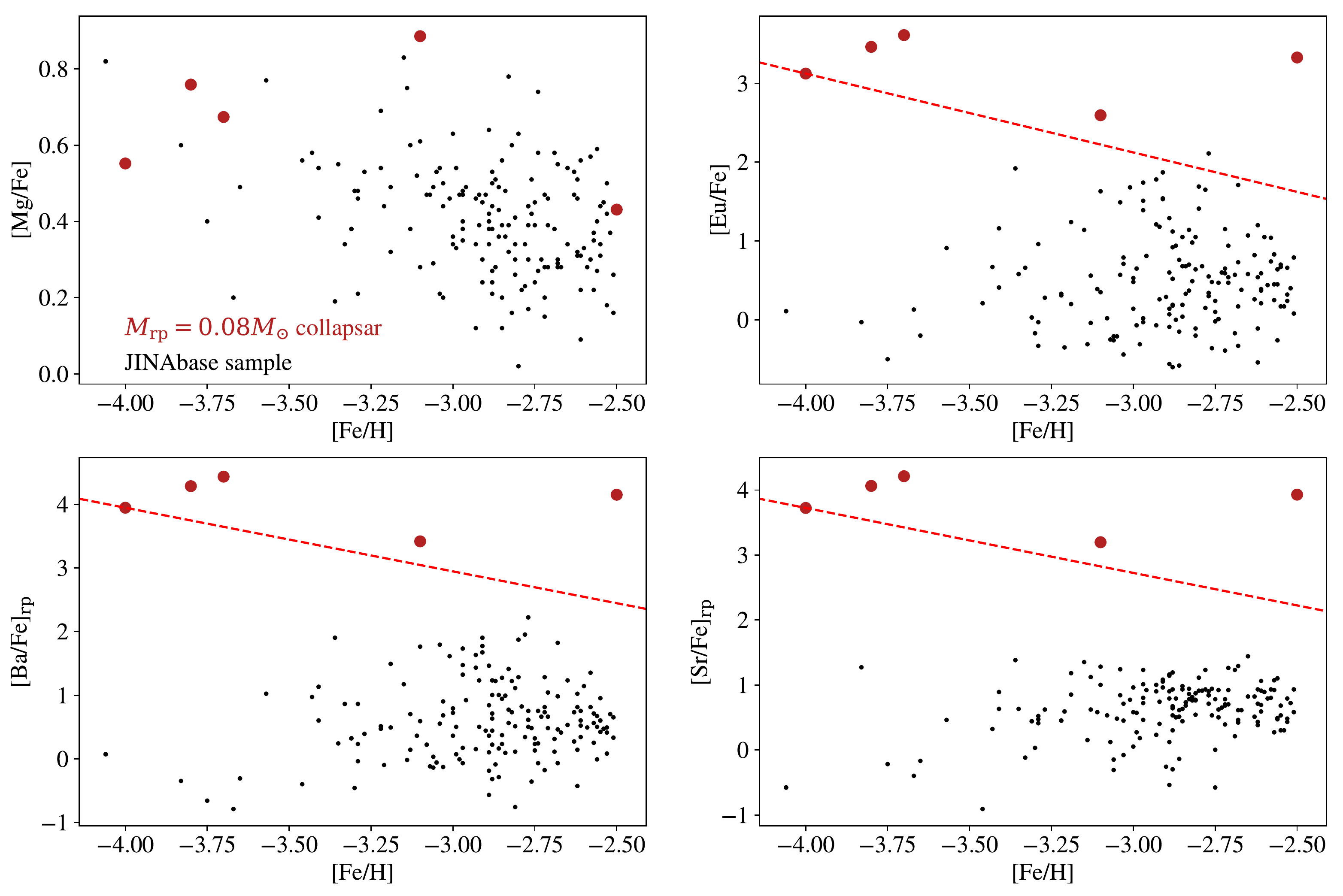}
\caption{ Stellar abundances from JINAbase are shown in black along with single-event predictions from the collapsar models of \citet{maeda2003} combined with \rp calculations from \citet{siegel2019b} in red. Average error bars are $\sim$ 0.2 dex for the data. While the alpha element Mg is in agreement with the data, the \rp abundances are much higher than seen in observations. Shown in the dashed line is the trajectory a parcel of gas would take due to dilution in Fe from nearby supernovae that provide Fe but no \rp elements. }
\label{fig:abundances}
\end{figure*}

In order to make predictions of [$r$/Fe] for our sample stars, we employ the nucleosynthesis calculations of \citet{maeda2003}.  They calculate bipolar (jetted) explosion models of 25 and 40 \msun\, stars while varying energy and opening angle as well as spherical ``normal" supernovae. We include results for only their bipolar explosions though the spherical results are qualitatively indistinguishable. We do not include their model 40D, which would have the largest discrepancy with the observational data due to an anomalously low Fe mass of $\sim10^{-7}$ \msun\, caused by a larger initial remnant mass.

These calculations provide us with Fe and Mg yields, as well as the [Fe/H] which depends on the mass over which the explosion diluted, given by 

\begin{equation}
M_{\rm dilution} = 5.1\times 10^{4} M_\odot E_{51}^{0.97}n_1^{-0.062}C_{10}^{-9/7},
\end{equation}
where $E_{51}$ is the explosion energy in units of $10^{51}$ ergs, $n_1$ is the ambient hydrogen number density in units of 1 cm$^{-3}$, and $C_{10}$ is the local sound speed in units of 10 km s$^{-1}$ \citep{shigeyama1998}. The Fe masses they obtain range between 0.078-0.54 \msun.  The calculations do not follow the nucleosynthesis in a remnant accretion disk and focus on the explosion itself, as the full calculation remains computationally prohibitive. Instead, we combine their calculations with the recent work by \citet{siegel2019b}, in which only the remnant accretion disk is simulated and \rp nucleosynthetic calculations are performed.

In order to derive a  mass of \rp elements necessarily synthesized in a single collapsar event to explain the Milky Way production rate, \citet{siegel2019b} convolve the long GRB and star formation rate with the average mass production of \rp elements in the Milky Way in order to derive a mass of $0.08-0.3$ \msun\, per event given the uncertainties. This is attainable within their simulations, thus introducing the viability of the collapsar model within their framework. In our analysis we adopt their lowest event yield of 0.08 \msun, thus deriving a lower bound on the expected [$r$/Fe] abundances.

After the collapsar event, the \rp material formed within the disk should efficiently mix with the ejecta synthesized in the explosion, thus leaving behind a chemical imprint on the subsequent generation of stars formed. With this, we are able to derive a simple equation to determine the expected [$r$/Fe] enhancement given by

\begin{equation} 
\label{eq:rfe} 
[r/{\rm Fe}] = {\rm log_{10}(}M_{\rm \text{r-p}}/M_{\rm Fe})_{\rm collapsar}-  {\rm log_{10}(}M_{\rm \text{r-p}}/M_{\rm Fe})_\odot,
\end{equation}
where $M_{\rm \text{r-p}}$ and $M_{\rm Fe}$ are the total mass of \rp elements and Fe produced in either the collapsar or contained within the solar abundance pattern. The solar abundance pattern is taken from \citet{lodders2003} and the \rp residual pattern from \citet{arnould2007}. For the solar \rp pattern, we consider elements starting from mass number A=69. 

This can be directly applied to any \rp element, assuming a solar abundance pattern is attained within the total \rp mass. The application of this equation to the lighter \rp elements may be less justified, as there is evidence of some variation within e.g. first-peak elements compared to the Sun. However its application to heavier elements is observationally motivated by the robustness of the heavy \rp solar pattern across decades in [Fe/H] metallicity \citep[e.g.,][]{frebel2018}. 

To test this prediction, we utilize JINAbase  \citep{abohalima2018} to gather stars with an [Fe/H] metallicity of < -2.5, from which measurements have also been made of Mg (alpha), Sr (first peak), Ba (second peak), and Eu (lanthanide). We employ an additional cut of [Ba/Eu] < -0.5 in order to avoid any early $s$-process contaminants \citep{simmerer2004}. We do not include stars with upper-limits on any of the elements we are considering. This brings our total sample to a size of 186 stars. At these low metallicities, stars formed should retain memory of at most a few nucleosynthetic events having polluted the gas from which they form \citep{audouze1995, shigeyama1998}.

With this we are able to directly compare the observations of heavy element abundances in our sample with predictions of the collapsar scenario. The results for Eu, Ba, and Sr are shown in Figure \ref{fig:abundances}. We compare against the solar \rp values for Sr and Ba by adjusting for the fact that the $s$-process contribution to the present day solar pattern is 75\% and 85\%, respectively. This only results in an offset in our plot as the correction is applied to both data and the models and does not inform any differences between them. As can be seen, the collapsar models consistently over-predict the observations of heavy elements. The top left panel shows the good agreement between predictions for Mg and our sample that is not seen in the \rp elements. We note that the largest discrepancies of $\sim$ 3 dex exist at the lowest metallicities, where the single-event assumption is most justified.

We assume that at large scales the ejecta mix efficiently, motivated by hydrodynamic simulations of the expansion of both collapsar jets and NSM tidal ejecta losing anisotropy early in their evolution \citep{ramirez2010,montes2016}. However, any geometric effects (e.g. disk winds being more focused along the midplane) would only serve to concentrate the \rp material, thus increasing the predicted [$r$/Fe] and exacerbating the discrepancy. Furthermore, though the $Y_{\rm e}$ electron fraction distribution may have a significant effect on limited to main \rp abundance ratios compared to Solar, the fact that the model predictions overshoot the data near all three \rp peaks does not allow for any reasonable abundance distribution to relieve the tension. For example, moving all of the \rp ejecta closer to the first peak, while conserving total mass, would only increase the predicted [Sr/Fe].

One possible resolution would be the dilution of [$r$/Fe] by typical core collapse supernovae that may not provide significant \rp enrichment. The red dashed line shows the result of such dilution. Since the Fe enhancement would only increase the denominator on the y-axis while increasing the numerator on the x-axis, this would result in the plotted line with slope -1 through the plot. This line does not intersect the data more than marginally for any star and is thus still difficult to reconcile with the observations. It may be possible however that the [$r$/Fe] of the gas is enhanced with Fe from many nearby supernovae, taking it downward along the red trajectory drawn in Figure \ref{fig:abundances}. Then, at [Fe/H] metallicity of $\sim$-2 is diluted in both Fe and Eu equally by pristine inflowing material, lowering the [Fe/H] and pushing it horizontally back toward the data.

Another possibility is continued expansion of an Eu and Fe-containing collapsar ejecta mixing significantly beyond $M_{\rm dilution}$. If material expands into ejecta from many nearby supernovae, it will not maintain the [Eu/Fe] characteristic of its own yields, but rather obtain an [Fe/H] (and thus [Eu/Fe]) corresponding to the average [Fe/H] of the combined sources. Although this would violate the long-held  and often utilized ``single-event" nature of these stars, which has been used to compare to lighter element abundances against SN models, we discuss the implications of this scenario below.

\subsection{Observational Implications within the Collapsar Model}

As mentioned above we can also ask the inverse question from our JINAbase sample. That is, given the inferred [Eu/Fe] and assuming an Eu mass of $M_{\rm Eu, \rm collapsar} = X^{\rm Eu}_{\text{r-p},\odot}M_{\text{r-p}, \rm collapsar} = 8\times10^{-5}$\msun, we can invert Equation \ref{eq:rfe} to solve for the mass of Fe that the Eu mixes with. Furthermore, we can then calculate the mass of pristine material that the total Fe must be spread over in order to maintain the low [Fe/H]. This method is independent of the models of \citet{maeda2003}, as we now take the [Eu/Fe] abundances from our sample as an input to the equation as opposed to just comparing against them.

\begin{figure}[h!]

\plotone{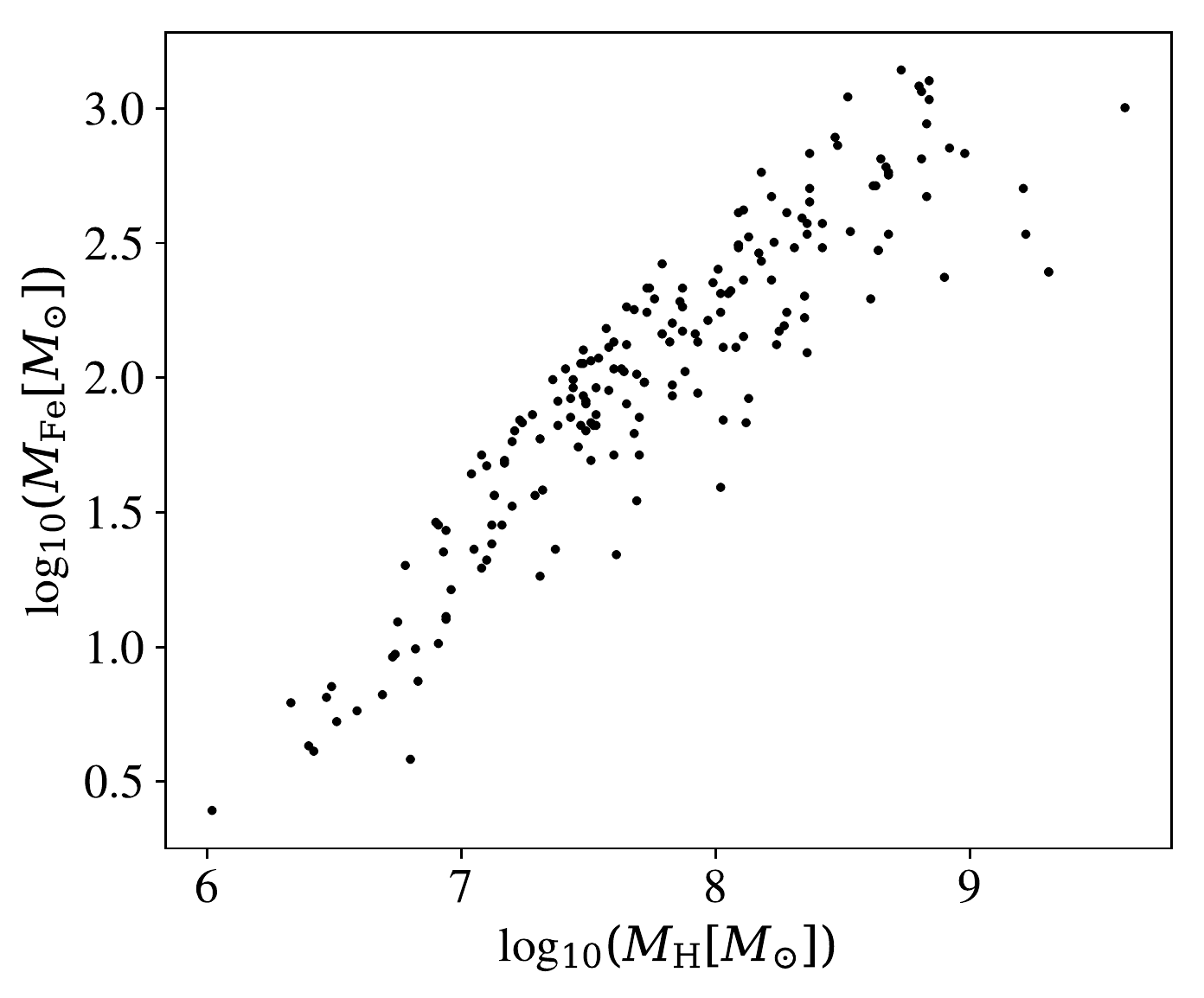}
\caption{Fe mass mixed with Eu producing events required to explain inferred [Eu/Fe] abundances are shown along the y-axis, and total Hydrogen mass in order to maintain the [Fe/H] in the stars is shown on the x-axis for our JINAbase sample of stars. }
\label{fig:fe_dilute}
\end{figure}

Figure \ref{fig:fe_dilute} shows the result of this experiment. We immediately see a $\sim 10^3$ dynamic range in both the inferred swept Fe masses and the Hydrogen dilution masses necessary to reproduce the observations. In the most extreme cases, we find Hydrogen dilution masses of $\approx 10^9$ \msun. This is well beyond the typical dilution masses of $\sim 10^4$ \msun, requiring a very large-scale mixing process which may be disfavored by the observed dispersion of \rp element abundances at early times, the rarity of the $r$-I/II stars ([Eu/Fe] > +0.3), as well as disputing the single-event nature of stars below our [Fe/H] < -2.5 selection.

We propose that a more likely resolution is that the \rp production site is spatially uncorrelated with any Fe production, and thus the \rp material is naturally diluted by the time it comes into contact with any Fe \citep{shen2015,vandevoort2015,naiman2018}. This may be the case for a NSM, in which substantial kicks during the preceding SNe may push NSM progenitors far from their birthplace before merging. While this argument does not solve the outstanding puzzle of the [Eu/Fe] ``knee" seen in the chemical evolution trend, it does argue against an \rp source synthesizing Fe, unless its mixing process is extreme.

\section{Discussion}\label{sec:discussion}

Our analysis finds that any \rp source that is physically associated with Fe production seems to be disfavored by observations of heavy elements within very metal-poor stars. While this finding is consistent with aspects of previous analyses \citep[c.f.,][and references therein]{qian2007}, we place direct constraints on the collapsar model of \rp enrichment. This poses a challenge to the collapsar scenario, instead lending further credence to a NSM or NS black-hole (NSBH) merger that should result in minimal Fe production and likely take place far from their birth sites. 

This does not rule out standard CCSNe as a source of a lighter \rp, as they may produce substantially less e.g.\ Sr per event given their rates and may still be consistent with the abundances. The abundances also are not in conflict with a magnetorotational jet-driven supernova \citep{winteler2012,nishimura2015}, as the mass of \rp is decreased by a factor of 10 and still consistent with most of the data in Figure \ref{fig:abundances} provided external Fe enhancement. However, this scenario may require very high ($\sim 10^{13}$ G) pre-collapse magnetic fields to produce elements such as Eu \citep{mosta2018}.

Recent kinematic studies of the highly \rp enhanced $r$-II stars show evidence of clustering in phase space \citep{roederer2018}, suggesting they may have been stripped from small dwarf galaxies like the highly \rp enhanced Reticulum II \citep{ji2016}. If this is the case, measurements of total \rp masses of say Eu may not be reliable, as the abundances may be driven by environmental effects such as low star formation efficiency in these galaxies as opposed to probing the event itself. However given their low gas mass, the amount of dilution shown in Figure \ref{fig:fe_dilute} necessary to maintain both the [Eu/Fe] and [Fe/H] may be more difficult to attain. We also note that observations of \rp enhanced stars are biased toward the halo and thus may be probing a primarily ex-situ population. Future observations of \rp enhanced bulge MW stars may distinguish if the production scenario is different between these two components.

Crucial to our analysis is the assumption that the \rp elements are well mixed with the Fe by the time the next generation of stars form. As the Fe is primarily produced in a jet and the \rp may be launched in a more equatorially concentrated disk outflow, it is possible that their different ejection speeds and times may inhibit efficient mixing, though again this may be in conflict with hydrodynamical simulations \citep{ramirez2010,montes2016}. As an example, \citet{martin2015} propose inefficient mixing between a polar and equatorial outflow as a possibility to explain the abundances seen in HD 122563, though they suppress the mixing between components by a factor of 50 as opposed to the $>10^3$ suppression required for our lowest metallicity stars. 

If this inefficient mixing scenario is what occurs, we are still able to state that the abundances in metal-poor stars observed argue against spatial correlation of \rp production with Fe, as this would result in a similar trajectory to those shown by the red dashed-lines in Figure \ref{fig:abundances}. This spatial correlation is to be expected in a collapsar that explodes promptly following a star-formation event and would not have time to migrate far from its birthplace. 

While SSS17a/AT2017gfo  served as the long sought after direct evidence of significant \rp production in NSMs, there remain many outstanding questions regarding their role throughout cosmic time. There is certainly still room if not evidence for multiple progenitor channels contributing significantly. Future observations of electromagnetic counterparts to gravitational wave sources will continue to elucidate remaining open questions in \rp nucleosynthesis and hopefully reveal new mysteries to explore. 

We thank the referee for comments which have improved the quality of this manuscript. We thank Stephan Rosswog and Charli Sakari for useful comments on an earlier version of this manuscript. This research has made use of NASA's Astrophysics Data System. P.M. is supported in part by the Heising-Simons Foundation. E.R.-R. is supported in part by David and Lucile Packard Foundation and the Niels Bohr Professorship from the DNRF. The UCSC team is supported in part by NASA grant NNG17PX03C, NSF grant AST-1518052, the Gordon \& Betty Moore Foundation, the Heising-Simons Foundation, and the David and Lucile Packard Foundation.

\software{ NumPy \citep{numpy};   matplotlib \citep{hunter2007}.}

\end{document}